\documentclass[prb,a4paper,twocolumn,floatfix,showpacs,showkeys,amsmath,amssymb,nobibnotes,altaffilletter]{revtex4-1}

\usepackage{graphicx}
\usepackage{pslatex}
\usepackage{xspace}
\usepackage{subfigure}

\newcommand{\degree}{$^{\circ}$}

\begin{document}

\preprint{}

\title{Energetics of Excited States in the Conjugated Polymer Poly(3-hexylthiophene)}

\author{Carsten Deibel}\email{deibel@physik.uni-wuerzburg.de}%
\author{Daniel Mack}%
\author{Julien Gorenflot}%
\affiliation{Experimental Physics VI, Julius-Maximilians-University of W{\"u}rzburg, 97074 W{\"u}rzburg, Germany}

\author{Achim Sch{\"o}ll}%
\author{Stefan Krause}%
\author{Friedrich Reinert}%
\affiliation{Gemeinschaftslabor f{\"u}r Nanoanalytik, Forschungszentrum Karlsruhe, 76021 Karlsruhe, and Julius-Maximilians-University of W{\"u}rzburg, 97074 W{\"u}rzburg, Germany}

\author{Daniel Rauh}%
\author{Vladimir Dyakonov}%
\affiliation{Experimental Physics VI, Julius-Maximilians-University of W{\"u}rzburg, 97074 W{\"u}rzburg, Germany}%
\affiliation{Bavarian Centre for Applied Energy Research (ZAE Bayern), 97074 W{\"u}rzburg, Germany}

\date{\today}

\begin{abstract}
There is an enormous potential in applying conjugated polymers in novel organic opto-electronic devices such as light emitting diodes and solar cells. Although prototypes and first products exist, a comprehensive understanding of the fundamental processes and energetics involved during photoexcitation is still lacking and limits further device optimisations. Here we report on a unique analysis of the excited states involved in charge generation by photoexcitation. On the model system poly(3-hexylthiophene) (P3HT), we demonstrate the general applicability of our novel approach. From photoemission spectroscopy of occupied and unoccupied states we determine the transport gap to 2.6~eV, which we show to be in agreement with the onset of photoconductivity by spectrally resolved photocurrent measurements. For photogenerated singlet exciton at the absorption edge, 0.7~eV of excess energy are required to overcome the binding energy; the intermediate charge transfer state is situated only 0.3~eV above the singlet exciton. Our results give direct evidence of energy levels involved in the photogeneration and charge transport within conjugated polymers.

\end{abstract}

\pacs{71.23.Cq;  79.60.Ht; 73.50.Pz; 78.55.Qr; 73.61.Ph}

\keywords{organic semiconductors; polymers; energy levels; photoelectron spectroscopy; photoluminescence; photoconductivity}

\maketitle

\section{Introduction}

In bulk heterojunction solar cells,\cite{so2008review,brabec2008book} which have already show about $6$\% power conversion efficiency,\cite{green2009review} the light is mainly absorbed in the conjugated polymer, generating singlet excitons, which are strongly bound due to the weak screening in organic semiconductors.\cite{hertel2008review} The exciton dissociation can be induced by acceptor molecules, often fullerene derivatives, and depends on the difference between the energy of the singlet exciton and the energy of the donor--acceptor polaron pair. If the latter state is energetically favourable, the excitons are dissociated by an ultrafast and very efficient electron transfer from the donor polymer.\cite{sariciftci1992} Unfortunately, the energy difference between exciton and polaron pair state is lost, leading to a lowering of the open circuit voltage, an important parameter determining the photovoltaic power conversion efficiency.\cite{ross2009,vandewal2009a} This example illustrates clearly why the energy levels of the utilized materials are of crucial importance for the fundamental understanding and, consequently, for the systematic optimization of organic solar cells.

The magnitude of the binding energies of excited states in organic semiconductors, in particular for polaron pairs, is a controversial topic.\cite{sariciftci1997book,hertel2008review} Lately, the notion that the binding energy of the primary excitation can be overcome by the thermal energy~\cite{moses2000} is becoming less accepted,\cite{hwang2008} being replaced by a consensus that binding energies are much larger than the thermal energy in conjugated polymers.\cite{hertel2008review} The exciton binding energy has been measured for a range of conjugated polymers, but the single particle gap has not been determined.\cite{liess1997,sakurai1997,vanderhorst2001} The reason for the non-comprehensive determination of energy levels in previous works is the difficulty to find a suitable, complementary choice of experimental techniques relevant to photoexcitation and the generation of free charges.

In this article we report on a novel approach which is capable of providing this data by applying several complementary experimental techniques. The investigation was performed systematically by photoelectron spectroscopy (PES), inverse photoelectron spectroscopy (IPES), optical absorption, field-induced photoluminescence quenching (PL($F$)), and wavelength sensitive external quantum efficiency measurements (EQE) on the same or comparable samples of the model system poly(3-hexylthiophene) (P3HT). By this unique combination, we can provide conclusive energy levels for the positive and negative polarons, the transport gap, the absorption gap, the exciton binding energy, as well as a value for the polaron pair binding energy.

\begin{figure}
	\includegraphics[width=8.0cm]{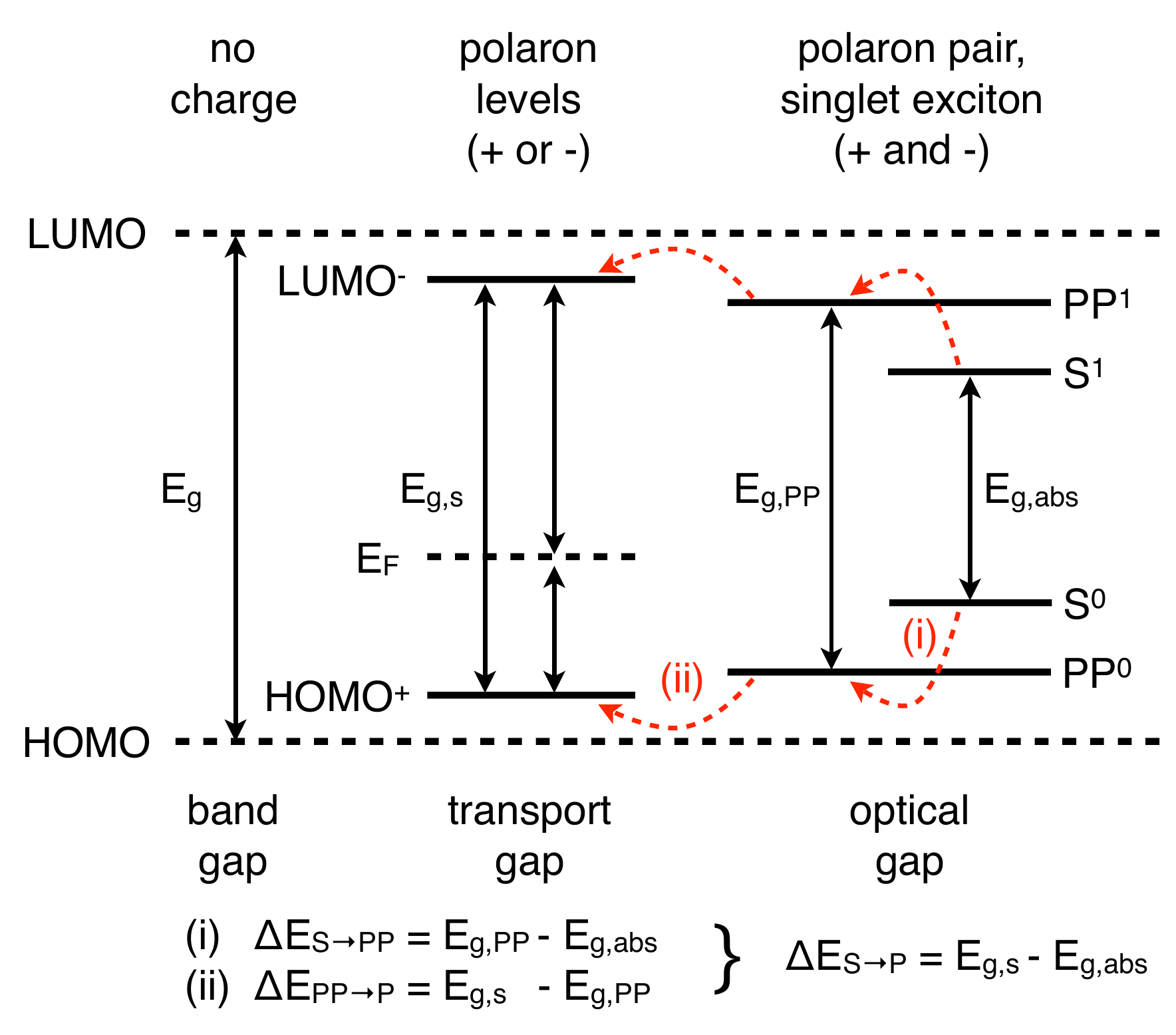}%
	\caption{(Color Online) Energy levels in a polymer in the ground state (a), during charge transport (b), and after photoexcitation (c). The details are described in the text.%
	\label{fig:E-levels}}
\end{figure}

Fig.~\ref{fig:E-levels} schematically shows the processes occuring during photoexcitation and charge transport for a typical polymeric semiconductor, as well as the respective energy levels. If charge is transported, the molecule will adapt to the (single) positive or negative charge in the HOMO or LUMO level, respectively, resulting in positive or negative polaronic levels (HOMO$^+$ and LUMO$^-$). They are separated by the transport gap, also called single particle gap $E_{g,s}$. In contrast, the absorption of a photon results in the generation of a correlated electron--hole pair, a singlet exciton. The separation of the respective excitonic levels $S^0$ and $S^1$ is the optical gap $E_{g,abs}$. During exciton dissociation, a strongly bound polaron pair (PP) with the levels $PP^0$ and $PP^1$ and the energy separation $E_{g,PP}$ can be formed. While the ground state band gap $E_g$ and the uncharged HOMO--LUMO levels can generally not be determined directly---particularly the common electrochemical methods being only sensitive to singly charged levels---a combination of techniques can shed light on the energetics involved in the generation and separation of charge.
The optical gap can be measured by optical absorption. The position of the polaronic transport levels (relative to the Fermi level $E_F$) and the transport gap can be accessed by photoelectron spectroscopy. In addition, photoluminescence quenching and external quantum efficiency measurements can provide the energies necessary to transfer the excitonic into bound or free polaron states, respectively.

\begin{figure}
   \includegraphics[width=8.0cm]{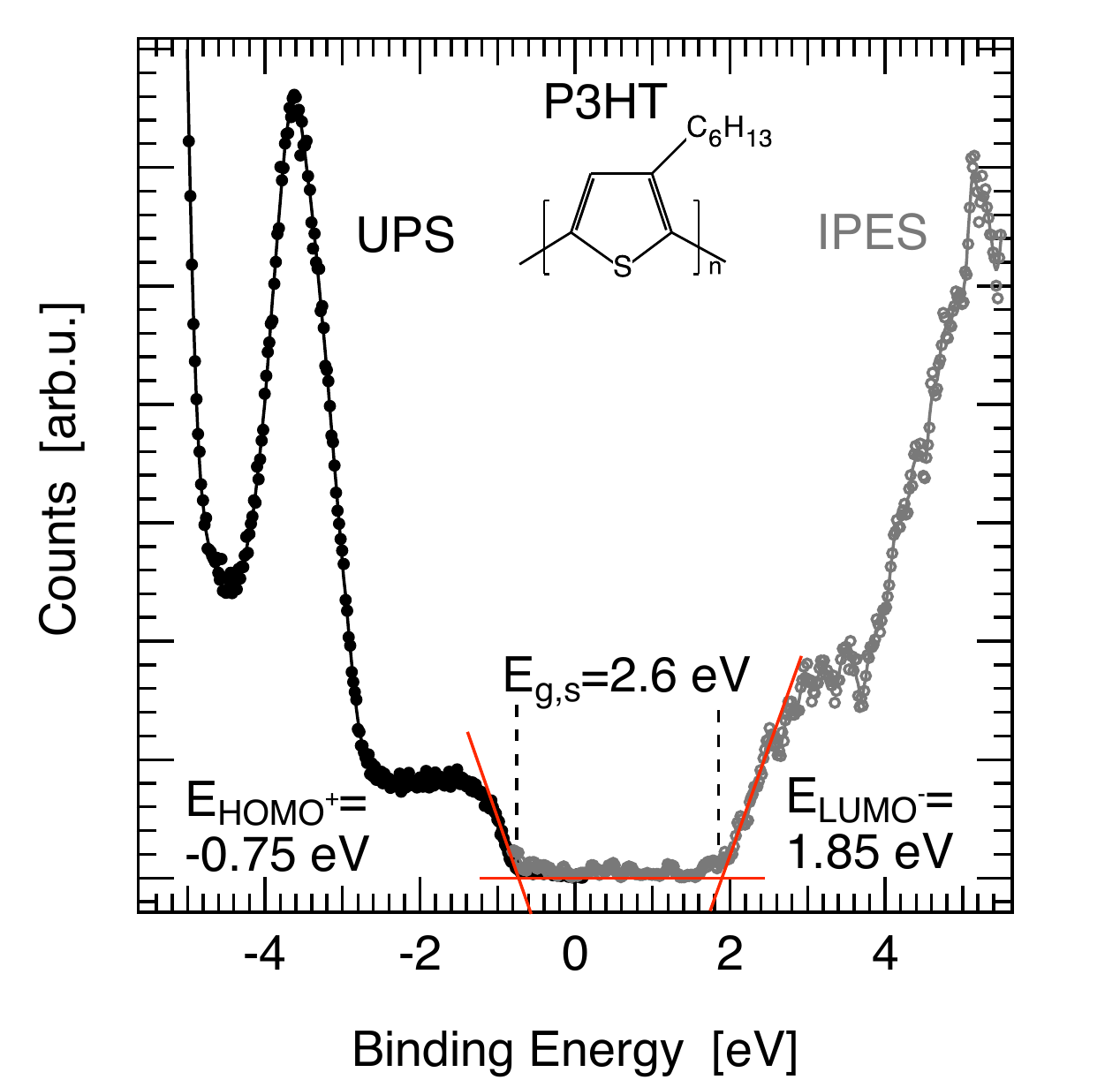}%
   \caption{(Color Online) UPS (black) and IPES (grey) spectra of a 70nm thick P3HT film on ITO/glass. Measured data are plotted as points, the straight lines are 5 point averages. The transport levels were derived from the band onsets as indicated by the lines. The transport gap is 2.6~eV. The molecular structure of P3HT is shown in the inset.%
   \label{fig:pes}}
\end{figure}


\section{Experimental Methods}

Thin films were processed by spin-coating a solution of 20~mg/ml of regioregular P3HT in chlorobenzene on oxygen-plasma treated indium tin oxide (ITO) coated glass substrates. The resulting layer thickness was 70~nm and 110~nm, as determined by a surface profiler (Veeco).
All samples were thermally treated at 120\degree{}C for 10 min. For diodes, 80~nm thick Al cathodes were thermally evaporated.
The P3HT was purchased from Rieke Metals.
According to the manufacturer, it has a regioregularity of between 90 and 93\%, and typically contains less than 0.08\% Ni, 0.5\% Zn and 0.7\% Br as residues from the synthesis.
For field-induced photoluminescence measurements, we additionally deposited two dielectric layers of 200~nm thickness by spincoating a solution of polymethylmethacrylat (PMMA), in order to avoid charge carrier injection. Thus, the device configuration was glass/ITO/(optional PMMA)/P3HT/(optional PMMA)/Al.

All UV photoelectron and inverse photoelectron spectroscopy measurements (UPS and IPES, respectively) were performed in a UHV system, described in detail elsewhere.\cite{eich2000} The base pressure was below $2\cdot10^{-10}$~mbar and the samples were at room temperature. The He-I line of a gas discharge lamp was used for UPS. The IPES detector (a Geiger--M{\"u}ller detector with SrF$_2$ window filled with Ar and I$_2$) was used in the isochromatic mode with a fixed photon energy of about 9.5~eV. The electron beam was defocussed over the sample to minimize radiation damage, and data acquisition was stopped immediately when changes in the IPES spectra occurred between two scans.

The field-induced photoluminescence was performed on samples held in a helium cold finger cryostat under vacuum. The excitation source was a mechanically modulated solid state laser at a wavelength 532~nm with a power of 30~mW. The photoluminescence was measured using a Cornerstone monochromator with a silicon diode and a liquid nitrogen cooled InSb-detector, and recorded by a lock-in amplifier. The electric field was applied by a Keithley 237 source measure unit. After each voltage step field and light were switched off for 4 minutes, which we found was the minimum time lag to avoid charging.

External quantum efficiency spectra were recorded by a homemade, lock-in based setup equipped with a calibrated reference silicon photodiode. A 300 W Xenon lamp was applied as light source.

\section{Experimental Results}

The combined UPS and IPES spectra of P3HT are presented in Fig.~\ref{fig:pes}. The band edges can be derived from the onset values of the HOMO and LUMO signals, as indicated by the lines used for evaluation.\cite{huefner1991} The resulting values, determined and given relative to the Fermi level $E_F$, are -0.75~eV for the HOMO and 1.85~eV for the LUMO. They can be associated with the respective relaxed ionic states, which correspond to the rearrangement of the electronic system due to the additional positive (negative) charge generated by (inverse) photoemission.\cite{krause2008,scholz2009a} In consequence, these values provide the position of the energy levels relevant for charge transport. The corresponding transport gap is the energetic separation between the respective transport levels and can be determined to 2.6~eV. Note that this is significantly different from other data in literature,\cite{feng2005} where the band energies are determined not from the onset values but the maxima of the signal peaks. 

\begin{figure}
   \includegraphics[width=8.0cm]{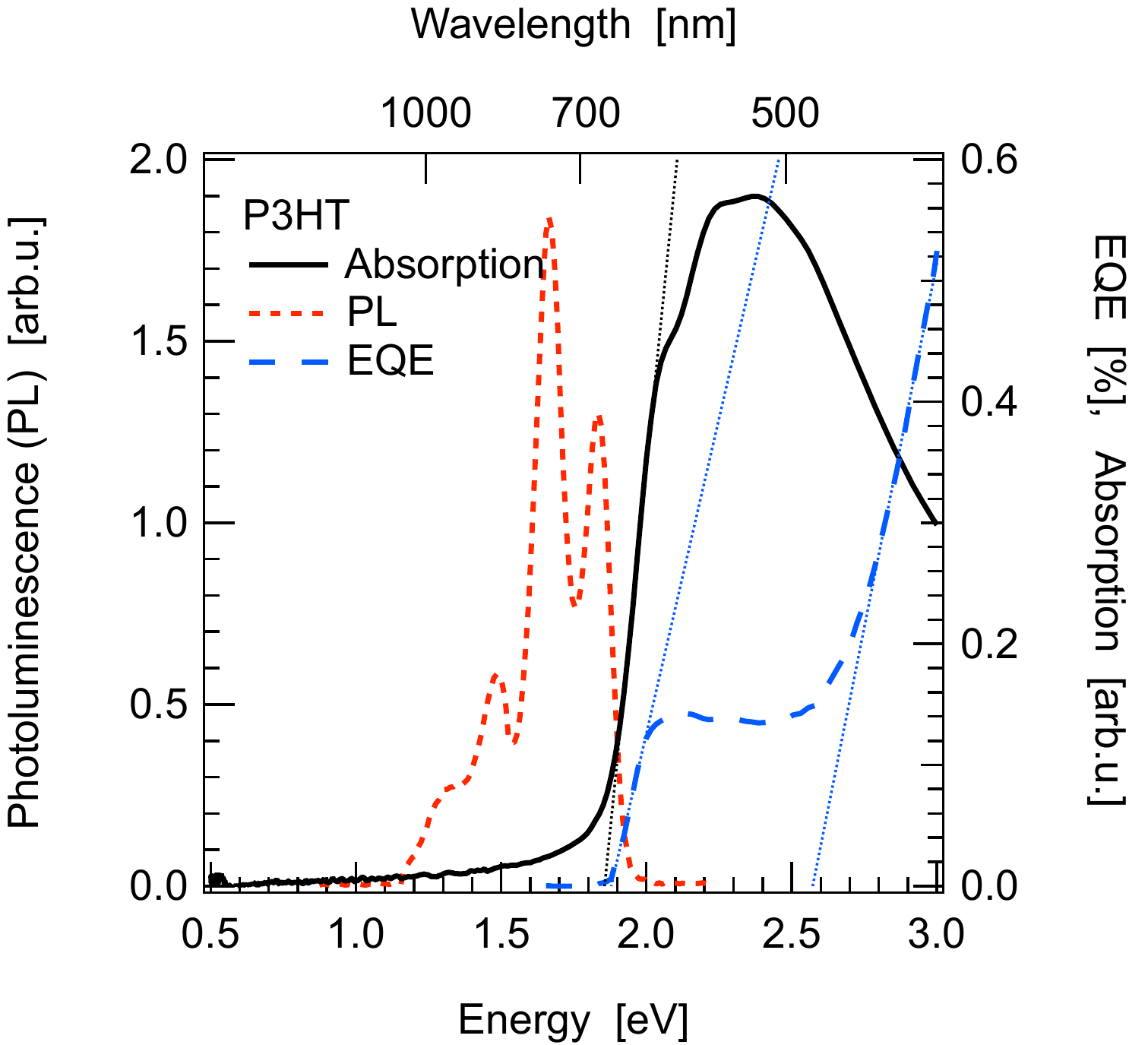}%
   \caption{(Color Online) Absorption (solid line), photoluminescence (dotted line) and external quantum efficiency (EQE) of P3HT thin films. The thin lines show how the onsets of the respective spectra were determined. The onset of the spectrally resolved photocurrent, the EQE, starts just above the absorption onset of 1.86~eV. The major photocurrent contribution, however, is seen at the energy of 2.58~eV, which corresponds to the transport gap extracted in Fig.~\ref{fig:pes}.%
   \label{fig:pl+abs}}
\end{figure}

In Fig.~\ref{fig:pl+abs} the absorption and photoluminescence spectra are shown for a 110 nm thick P3HT film on ITO. The absorption onset is at 1.85~eV and the maximum of the first singlet exciton transition $^1B_u$ is at about 2.1~eV. By applying a Gaussian fit to the highest energy photoluminescence peak at about 1.85~eV, we can determine the line width to $\sigma=68$~meV. This distribution also describes the absorption edge very well, and is assigned to the variance of the gaussian density of states distribution. The Stokes shift between the 0--0 absorption and emission band is about 250~meV.

These measurements are complemented by EQE spectra. This experimental technique shows the spectrally resolved photon-conversion yield, i.e., the fraction of generated and extracted charges per incident photon in dependence on the photon energy. The low energy region of such an EQE spectrum of the P3HT sample is presented in Fig.~\ref{fig:pl+abs}, measured under short circuit conditions. It shows a clear---although weak---onset of the photocurrent already at the absorption edge of 1.85~eV, with a quantum efficiency of about $0.1\%$. This implies that a fraction of the singlet excitons is already dissociated without excess energy. Synthesis residues such as Ni---typically below 100 parts per million relative to P3HT---or dissociation at the electrodes can be responsible for this low energy photocurrent, and would explain its low magnitude. Apart from this relatively weak signal, which is thus probably due to extrinsic effects,  the photocurrent strongly increases at photon energies above approx. 2.6~eV, and thus equals the intrinsic transport gap as determined by our photoemission experiments.

\begin{figure}
   \includegraphics[width=8.0cm]{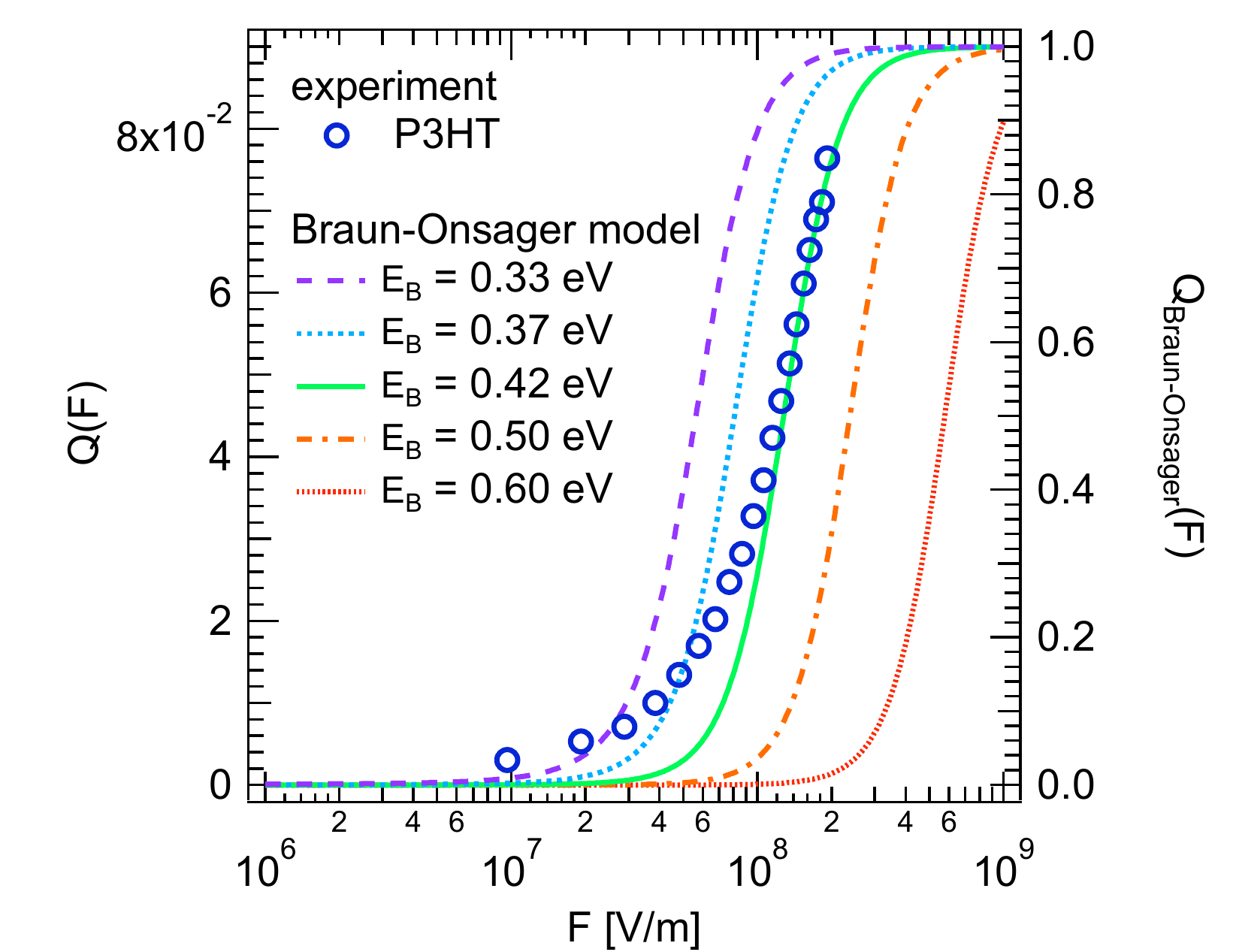}%
   \caption{(Color Online) The photoluminescence quenching yield $Q(F)=(PL(0)-PL(F))/PL(0)$ of a P3HT diode with dielectric layers. At high fields, it becomes apparent that not all luminescing singlet excitons can be quenched by the fields attainable with the experimental setup.%
   \label{fig:pl-field}}
\end{figure}

The electric field dependence of the P3HT luminescence peak of 1.65~eV is shown in Fig.~\ref{fig:pl-field}. The spectral shape of the PL is field independent, only the magnitude changes (not shown). The field-induced photoluminescence is a suitable technique to determine the P3HT singlet exciton binding energy.  We point out that the magnitude of the binding energy cannot be extracted directly. Therefore, we related our field-dependent PL quenching measurements to the established Onsager--Braun\cite{onsager1938,braun1984} (OB) theory for calculating the separation yield of the bound state, which considers the finite lifetime of the excited state. Although this empirical theory~\cite{wojcik2009} was originally derived for ion pair dissociation, is has been found suitable for calculating the exciton binding energy.\cite{sariciftci1997book,hertel2008review}
Within this framework, the dissociation probability $P(F)$ is given by\cite{deibel2009a}
\begin{eqnarray}
   P(F) & = & \frac{k_d(F)}{k_d(F) + k_f} = \frac{\kappa_d(F)}{\kappa_d(F) + (\mu \tau_f)^{-1}} .
   \label{eqn:P}
\end{eqnarray}
$F$ is the electric field, $\tau_f=k_f^{-1}$ the exciton lifetime, $\mu$ the sum of electron and hole mobilities. The field dependent dissociation rate $k_d(F)=\mu \kappa_d(F)$ is
\begin{equation}
   k_d(F) = \frac{3\gamma}{4\pi r_{S}^3} \exp\left( - \frac{E_b}{kT} \right) \frac{J_1\left( 2 \sqrt{-2b}  \right)}{\sqrt{-2b}}
   \label{eqn:kd}
\end{equation}
where $\gamma=q\mu/\epsilon\epsilon_0$ is the Langevin recombination factor,\cite{langevin1903} $r_S$ is the initial exciton radius, $E_b=e^2 / (4\pi\epsilon\epsilon_0 r_S)$ is the exciton binding energy, $kT$ the thermal energy, $J_1$ the Bessel function of order one, and $b=e^3F/(8\pi\epsilon\epsilon_0(kT)^2)$. $F$ is the electric field, $e$ is the elementary charge, and $\epsilon\epsilon_0$ the effective dielectric constant of the organic semiconductor blend.

Using the OB model with $\mu\tau=5\cdot10^{-16}$~m$^2$/V, $\epsilon=3.4$ and a maximum quenching of 9\%, we could achieve a reasonable fit to our experiment, yielding $E_b=0.42$~eV as shown in Fig.~\ref{fig:pl-field}. The discrepancy between the shoulder at low fields in the experimental data and the fit can be explained by the reasonable assumption that the exciton binding energies are distributed over a finite energy range. We attribute the different magnitudes of experimental and modelled the field dependent photoluminescence quenching to higher energy transistions from singlet exciton to polaron pair, which are beyond our measurement range.

Using the Einstein relation, the $\mu\tau$ product can also be expressed in terms of a diffusion length $L_D$,
\begin{equation}
   L_D=\sqrt{D\tau}=\sqrt{\frac{kT}{e} \mu \tau},
\end{equation}
$D$ being the diffusion coefficient. For $\mu\tau=5\cdot10^{-16}$~m$^2$/V, $L_D$ becomes $3.5$~nm. Using the singlet exciton lifetime as determined recently for regioregular P3HT,\cite{shaw2008} $\tau=0.4$~ns, we determine a mobility of the strongly bound charges of about $1.3\cdot 10^{-6}$~m$^2/$Vs from the above mentioned $\mu\tau$ product used to describe the data in Fig.~\ref{fig:pl-field}.


\section{Discussion}

\begin{table}[htdp]
   \caption{Experimentally determined energy levels of regioregular P3HT. The corresponding energies are shown and explained in Fig.~\ref{fig:E-levels}.}
       \begin{center}
           \begin{tabular}{l@{\hspace{0.35cm}}r@{\hspace{0.35cm}}l}
               Energy              & [eV]  & Method\\ \hline\hline
               $E_{g,s}$               & $2.6$ & PES/IPES\\
               LUMO$^-$ relative to $E_F$  & $1.85$    & PES/IPES \\
               HOMO$^+$ relative to $E_F$  & $-0.75$   & PES/IPES \\\hline
               $E_{g,abs}$         & $1.85$    & absorption \\ \hline
               $E_{g,s}$               & $2.58$    & EQE \\
               $E_{g,abs}$         & $1.88$    & EQE \\ \hline
               $\Delta E_{S\rightarrow PP}$    &$ \ge 0.42$    & field dependent PL \\
               $\Delta E_{S\rightarrow P}$ & $\approx 0.7$ & PES/IPES vs.\ absorption, EQE\\
           \end{tabular}
       \end{center}
   \label{tab:E-levels}
\end{table}%

The experimentally determined energy levels of regioregular P3HT, schematically shown and explained in Fig.~\ref{fig:E-levels}, are summarized in Tab.~\ref{tab:E-levels}. A central result is the finding that the transport gap $E_{g,s}$ of 2.6~eV directly corresponds to the energy needed for achieving an efficient photocurrent. We point out that the two independent methods involved in determining this single particle gap---photoemission spectroscopy and external quantum efficiency---rely on very different principles. The former is very surface sensitive and is performed on thin films, whereas the latter relies on charge generation and transport in the bulk of a device. The nevertheless same resulting energy for the transport gap is consistent with former work.\cite{casu2007} The difference between $E_{g,s}$ and the absorption onset $E_{g,abs}=1.85$~eV can be attributed to the exciton binding energy $\Delta E_{S\rightarrow P} \simeq 0.7$~eV. Sakurai et al.\cite{sakurai1997} investigated  poly(3-octylthiophene) and found the  $^1A_g$ polaron pair state 0.5~eV above the $^1B_u$ exciton by electroabsorption and two-photon absorption. Using electroabsorption, Liess et al.\cite{liess1997} determined 0.6~eV for the transition from the $^1B_u$ exciton state to the $^mA_g$ state of another poly(3-alkyl thiophene), pointing out that this might be lower limit for the continuum band threshold. Van der Horst \emph{et al.} calculated 0.61~eV for the latter transition.\cite{vanderhorst2001} We note that our experimental approach, in contrast to Sakurai \emph{et al.}, Liess \emph{et al.} and van der Horst \emph{et al.}, probes the transition from singlet excitons to free polarons, thus giving the exciton binding energy relevant for applications such as photovoltaics.

The apparently lower exciton binding energy found by PL($F$) measurements, $0.42$~eV instead of $0.7$~eV, can be understood as follows. In order to quench the PL, it is sufficient to separate the constituents of the exciton partly, i.e., generating a---still bound---polaron pair. Therefore, not the full energy to generate a free polaron from the exciton has to be invested. In other words, the PL quenching experiment determines the energy of the transition from singlet exciton to polaron pair, $\Delta E_{S\rightarrow PP} \ge 0.42$~eV. Note that from the levelling off of the field dependent PL shown in Fig.~\ref{fig:pl-field} it is clear that only a fraction of the singlet exciton radiative decay is suppressed. This effect is probably related to the semicrystalline regions in P3HT,\cite{sirringhaus1999} leading to self-quenching of PL even at zero field. Also, the exciton binding energy, and specifically the exciton to polaron pair transition, is probably distributed in energy. The experimentally determined value of 0.42~eV thus presents a lower limit for the exciton-to-polaron-pair transition energy. It corresponds well with the density functional theory calculations of van der Horst~\cite{vanderhorst2001} for polythiophene, where 0.45~eV were determined for this transition, and the photophysical experiments of Sakurai et al.\cite{sakurai1997} on poly(3-alkyl thiophene) (0.55~eV).

The remaining energy difference between polaron pair and free polaron, which is necessary to create a photocurrent, can be expressed as $\Delta E_{S\rightarrow P}-\Delta E_{S\rightarrow PP}$. Experimentally, we find that $\le 0.3$~eV are needed for this transition. This energy difference corresponds to a Coulomb binding energy. Assuming a dielectric constant $\epsilon=3.4$, an electron--hole distance of about 1.4~nm can be derived, which is similar to the d-spacing of 1.6~nm found in X-ray diffraction.\cite{vanlaeke2006a} The latter can be attributed to twice the hexyl side chain length of P3HT, representing the stacking of different molecular units.

We point out that our results suggest that the interpretation of {\"O}sterbacka et al.\cite{osterbacka2000}  concerning polaron transitions observed in P3HT by photoinduced absorption spectroscopy might need to be reconsidered. In Ref.~\cite{osterbacka2000}, two peaks assigned to localized polarons, P$_1$ and P$_2$, are reported to absorb at 0.45~eV and 1.3~eV in regioregular and regiorandom P3HT. Two other peaks assigned to delocalized polarons, DP$_1$ and DP$_2$, are only observed in regioregular P3HT at 60~meV resp.\ 1.8~eV. {\"O}sterbacka et al. explain their findings on basis of charged HOMO and LUMO levels moving closer together into the HOMO--LUMO band gap due to the polaronic relaxation. The two polaron signals are then explained by optical transitions from the HOMO to the charged HOMO, corresponding to P$_1$, and from there to the charged LUMO, corresponding to P$_2$. In this interpretation, the HOMO--LUMO gap should correspond to 2$\times$P$_1+$P$_2$=2.1~eV resp.\ to 2$\times$DP$_1+$DP$_2$=1.92~eV. Both these energy differences are much smaller than the transport gap---which we determined to be 2.6~eV with two independent techniques---let alone the HOMO--LUMO gap. However, on basis of our data we currently cannot present a concise alternative explanation.


\section{Conclusions}

In conclusion, we have presented an experimental approach, which is capable of providing the position of all energy levels relevant during photoexcitation and charge transport within a conjugated polymer. For the first time, we determine the single particle gap of the model polymer poly(3-hexylthiophene) (P3HT) relevant for charge transport to 2.6~eV by PES/IPES measurements, and show with EQE experiments that it is equivalent to the onset of photogeneration. By comparing this data to optical absorption, the exciton binding energy is determined for regioregular P3HT to about 0.7~eV. Moreover, from a comparison to field dependent photoluminescence measurements, we can provide a lower limit of 0.42~eV for the transition from a singlet exciton to a bound polaron pair and a polaron pair binding energy of about 0.3~eV. As the method of field induced photoluminescence quenching was used commonly to determine singlet exciton binding energies, but actually yields the polaron pair binding energy, previous experimental results should be reconsidered. All in all, we provide a comprehensive insight into the energy levels and binding energies of excited states in the frequently used compound P3HT. 
We emphasize that the presented experimental approach is applicable to conjugated polymers in general, and will be used to expand our findings in view of a comprehensive understanding of this class of materials.

\section*{Acknowledgments}

Discussions with K.~Kanai are acknowledged. The current work is supported by the Bundesministerium f{\"u}r Bildung und Forschung in the framework of the GREKOS project (contract no.~03SF0356B) and by the DFG (GK 1221). J.G.'s work is funded by the solarNtype project (contract no.~035533-2) within the 6th framework programme of the EU. V.D.'s work at the ZAE Bayern is financed by the Bavarian Ministry of Economic Affairs, Infrastructure, Transport and Technology.


\begin{thebibliography}{10}%
\makeatletter
\providecommand \@ifxundefined [1]{%
 \ifx #1\undefined \expandafter \@firstoftwo
 \else \expandafter \@secondoftwo
\fi
}%
\providecommand \@ifnum [1]{%
 \ifnum #1\expandafter \@firstoftwo
 \else \expandafter \@secondoftwo
\fi
}%
\providecommand \enquote [1]{``#1''}%
\providecommand \bibnamefont  [1]{#1}%
\providecommand \bibfnamefont [1]{#1}%
\providecommand \citenamefont [1]{#1}%
\providecommand\href[0]{\@sanitize\@href}%
\providecommand\@href[1]{\endgroup\@@startlink{#1}\endgroup\@@href}%
\providecommand\@@href[1]{#1\@@endlink}%
\providecommand \@sanitize [0]{\begingroup\catcode`\&12\catcode`\#12\relax}%
\@ifxundefined \pdfoutput {\@firstoftwo}{%
 \@ifnum{\z@=\pdfoutput}{\@firstoftwo}{\@secondoftwo}%
}{%
 \providecommand\@@startlink[1]{\leavevmode}%
 \providecommand\@@endlink[0]{}%
}{%
 \providecommand\@@startlink[1]{%
  \leavevmode
  \pdfstartlink
   attr{/Border[0 0 1 ]/H/I/C[0 1 1]}%
   user{/Subtype/Link/A<</Type/Action/S/URI/URI(#1)>>}%
  \relax
 }%
 \providecommand\@@endlink[0]{\pdfendlink}%
}%
\providecommand \url  [0]{\begingroup\@sanitize \@url }%
\providecommand \@url [1]{\endgroup\@href {#1}{\urlprefix}}%
\providecommand \urlprefix [0]{URL }%
\providecommand \Eprint[0]{\href }%
\@ifxundefined \urlstyle {%
  \providecommand \doi [1]{doi:\discretionary{}{}{}#1}%
}{%
  \providecommand \doi [0]{doi:\discretionary{}{}{}\begingroup
  \urlstyle{rm}\Url }%
}%
\providecommand \doibase [0]{http://dx.doi.org/}%
\providecommand \Doi[1]{\href{\doibase#1}}%
\providecommand \bibAnnote [3]{%
  \BibitemShut{#1}%
  \begin{quotation}\noindent
    \textsc{Key:}\ #2\\\textsc{Annotation:}\ #3%
  \end{quotation}%
}%
\providecommand \bibAnnoteFile [2]{%
  \IfFileExists{#2}{\bibAnnote {#1} {#2} {\input{#2}}}{}%
}%
\providecommand \typeout [0]{\immediate \write \m@ne }%
\providecommand \selectlanguage [0]{\@gobble}%
\providecommand \bibinfo [0]{\@secondoftwo}%
\providecommand \bibfield [0]{\@secondoftwo}%
\providecommand \translation [1]{[#1]}%
\providecommand \BibitemOpen[0]{}%
\providecommand \bibitemStop [0]{}%
\providecommand \bibitemNoStop [0]{.\EOS\space}%
\providecommand \EOS [0]{\spacefactor3000\relax}%
\providecommand \BibitemShut [1]{\csname bibitem#1\endcsname}%
\bibitem{so2008review}%
  \BibitemOpen
  \bibfield{author}{%
  \bibinfo {author} {\bibfnamefont{F.}~\bibnamefont{So}}, \bibinfo {author}
  {\bibfnamefont{J.}~\bibnamefont{Kido}},\ and\ \bibinfo {author}
  {\bibfnamefont{P.}~\bibnamefont{Burrows}},\ }%
  \bibfield{journal}{%
  \bibinfo {journal} {MRS Bulletin}\ }%
  \textbf{\bibinfo {volume} {33}},\ \bibinfo {pages} {663} (\bibinfo {year}
  {2008})%
  \bibAnnoteFile{NoStop}{so2008review}%
\bibitem{brabec2008book}%
  \BibitemOpen
  \bibfield{author}{%
  \bibinfo {author} {\bibfnamefont{C.}~\bibnamefont{Brabec}}, \bibinfo {author}
  {\bibfnamefont{U.}~\bibnamefont{Scherf}},\ and\ \bibinfo {author}
  {\bibfnamefont{V.}~\bibnamefont{Dyakonov}},\ }%
  \emph{\bibinfo {title} {Organic Photovoltaics}}\ (\bibinfo {publisher} {Wiley
  VCH},\ \bibinfo {address} {Weinheim, Germany},\ \bibinfo {year} {2008})%
  \bibAnnoteFile{NoStop}{brabec2008book}%
\bibitem{green2009review}%
  \BibitemOpen
  \bibfield{author}{%
  \bibinfo {author} {\bibfnamefont{M.~A.}\ \bibnamefont{Green}}, \bibinfo
  {author} {\bibfnamefont{K.}~\bibnamefont{Emery}}, \bibinfo {author}
  {\bibfnamefont{Y.}~\bibnamefont{Hishikawa}},\ and\ \bibinfo {author}
  {\bibfnamefont{W.}~\bibnamefont{Warta}},\ }%
  \bibfield{journal}{%
  \bibinfo {journal} {Prog. Photovolt.}\ }%
  \textbf{\bibinfo {volume} {17}},\ \bibinfo {pages} {85} (\bibinfo {year}
  {2009})%
  \bibAnnoteFile{NoStop}{green2009review}%
\bibitem{hertel2008review}%
  \BibitemOpen
  \bibfield{author}{%
  \bibinfo {author} {\bibfnamefont{D.}~\bibnamefont{Hertel}}\ and\ \bibinfo
  {author} {\bibfnamefont{H.}~\bibnamefont{B{\"a}ssler}},\ }%
  \bibfield{journal}{%
  \bibinfo {journal} {Chem. Phys. Chem.}\ }%
  \textbf{\bibinfo {volume} {9}},\ \bibinfo {pages} {666} (\bibinfo {year}
  {2008})%
  \bibAnnoteFile{NoStop}{hertel2008review}%
\bibitem{sariciftci1992}%
  \BibitemOpen
  \bibfield{author}{%
  \bibinfo {author} {\bibfnamefont{N.~S.}\ \bibnamefont{Sariciftci}}, \bibinfo
  {author} {\bibfnamefont{L.}~\bibnamefont{Smilowitz}}, \bibinfo {author}
  {\bibfnamefont{A.~J.}\ \bibnamefont{Heeger}},\ and\ \bibinfo {author}
  {\bibfnamefont{F.}~\bibnamefont{Wudl}},\ }%
  \bibfield{journal}{%
  \bibinfo {journal} {Science}\ }%
  \textbf{\bibinfo {volume} {258}},\ \bibinfo {pages} {1474} (\bibinfo {year}
  {1992})%
  \bibAnnoteFile{NoStop}{sariciftci1992}%
\bibitem{ross2009}%
  \BibitemOpen
  \bibfield{author}{%
  \bibinfo {author} {\bibfnamefont{R.~B.}\ \bibnamefont{Ross}}, \bibinfo
  {author} {\bibfnamefont{C.~M.}\ \bibnamefont{Cardona}}, \bibinfo {author}
  {\bibfnamefont{D.~M.}\ \bibnamefont{Guldi}}, \bibinfo {author}
  {\bibfnamefont{S.~G.}\ \bibnamefont{Sankaranarayanan}}, \bibinfo {author}
  {\bibfnamefont{M.~O.}\ \bibnamefont{Reese}}, \bibinfo {author}
  {\bibfnamefont{N.}~\bibnamefont{Kopidakis}}, \bibinfo {author}
  {\bibfnamefont{J.}~\bibnamefont{Peet}}, \bibinfo {author}
  {\bibfnamefont{B.}~\bibnamefont{Walker}}, \bibinfo {author}
  {\bibfnamefont{G.~C.}\ \bibnamefont{Bazan}}, \bibinfo {author}
  {\bibfnamefont{E.~V.}\ \bibnamefont{Keuren}}, \bibinfo {author}
  {\bibfnamefont{B.~C.}\ \bibnamefont{Holloway}},\ and\ \bibinfo {author}
  {\bibfnamefont{M.}~\bibnamefont{Drees}},\ }%
  \bibfield{journal}{%
  \bibinfo {journal} {Nat. Mater.}\ }%
  \textbf{\bibinfo {volume} {8}},\ \bibinfo {pages} {208} (\bibinfo {year}
  {2009})%
  \bibAnnoteFile{NoStop}{ross2009}%
\bibitem{vandewal2009a}%
  \BibitemOpen
  \bibfield{author}{%
  \bibinfo {author} {\bibfnamefont{K.}~\bibnamefont{Vandewal}}, \bibinfo
  {author} {\bibfnamefont{K.}~\bibnamefont{Tvingstedt}}, \bibinfo {author}
  {\bibfnamefont{A.}~\bibnamefont{Gadisa}}, \bibinfo {author}
  {\bibfnamefont{O.}~\bibnamefont{Ingan{\"a}s}},\ and\ \bibinfo {author}
  {\bibfnamefont{J.~V.}\ \bibnamefont{Manca}},\ }%
  \bibfield{journal}{%
  \bibinfo {journal} {Nat. Mater.}\ }%
  \textbf{\bibinfo {volume} {8}},\ \bibinfo {pages} {904} (\bibinfo {year}
  {2009})%
  \bibAnnoteFile{NoStop}{vandewal2009a}%
\bibitem{sariciftci1997book}%
  \BibitemOpen
  \emph{\bibinfo {title} {The nature of primary photoexcitations in conjugated
  polymers}},\ edited by\ \bibinfo {editor} {\bibfnamefont{N.~S.}\
  \bibnamefont{Sariciftci}}\ (\bibinfo {publisher} {World Scientific},\
  \bibinfo {address} {Singapore},\ \bibinfo {year} {1997})%
  \bibAnnoteFile{NoStop}{sariciftci1997book}%
\bibitem{moses2000}%
  \BibitemOpen
  \bibfield{author}{%
  \bibinfo {author} {\bibfnamefont{D.}~\bibnamefont{Moses}}, \bibinfo {author}
  {\bibfnamefont{A.}~\bibnamefont{Dogariu}},\ and\ \bibinfo {author}
  {\bibfnamefont{A.~J.}\ \bibnamefont{Heeger}},\ }%
  \bibfield{journal}{%
  \bibinfo {journal} {Phys. Rev. B}\ }%
  \textbf{\bibinfo {volume} {61}},\ \bibinfo {pages} {9373} (\bibinfo {year}
  {2000})%
  \bibAnnoteFile{NoStop}{moses2000}%
\bibitem{hwang2008}%
  \BibitemOpen
  \bibfield{author}{%
  \bibinfo {author} {\bibfnamefont{I.-W.}\ \bibnamefont{Hwang}}, \bibinfo
  {author} {\bibfnamefont{D.}~\bibnamefont{Moses}},\ and\ \bibinfo {author}
  {\bibfnamefont{A.~J.}\ \bibnamefont{Heeger}},\ }%
  \bibfield{journal}{%
  \Doi{10.1021/jp075565x}{\bibinfo {journal} {J. Phys. Chem. C}}\ }%
  \textbf{\bibinfo {volume} {112}},\ \bibinfo {pages} {4350} (\bibinfo {year}
  {2008})%
  \bibAnnoteFile{NoStop}{hwang2008}%
\bibitem{liess1997}%
  \BibitemOpen
  \bibfield{author}{%
  \bibinfo {author} {\bibfnamefont{M.}~\bibnamefont{Liess}}, \bibinfo {author}
  {\bibfnamefont{S.}~\bibnamefont{Jeglinski}}, \bibinfo {author}
  {\bibfnamefont{Z.~V.}\ \bibnamefont{Vardeny}}, \bibinfo {author}
  {\bibfnamefont{M.}~\bibnamefont{Ozaki}}, \bibinfo {author}
  {\bibfnamefont{K.}~\bibnamefont{Yoshino}}, \bibinfo {author}
  {\bibfnamefont{Y.}~\bibnamefont{Ding}},\ and\ \bibinfo {author}
  {\bibfnamefont{T.}~\bibnamefont{Barton}},\ }%
  \bibfield{journal}{%
  \bibinfo {journal} {Phys. Rev. B}\ }%
  \textbf{\bibinfo {volume} {56}},\ \bibinfo {pages} {15712} (\bibinfo {year}
  {1997})%
  \bibAnnoteFile{NoStop}{liess1997}%
\bibitem{sakurai1997}%
  \BibitemOpen
  \bibfield{author}{%
  \bibinfo {author} {\bibfnamefont{K.}~\bibnamefont{Sakurai}}, \bibinfo
  {author} {\bibfnamefont{H.}~\bibnamefont{Tachibana}}, \bibinfo {author}
  {\bibfnamefont{N.}~\bibnamefont{Shiga}}, \bibinfo {author}
  {\bibfnamefont{C.}~\bibnamefont{Terakura}}, \bibinfo {author}
  {\bibfnamefont{M.}~\bibnamefont{Matsumoto}},\ and\ \bibinfo {author}
  {\bibfnamefont{Y.}~\bibnamefont{Tokura}},\ }%
  \bibfield{journal}{%
  \bibinfo {journal} {Phys. Rev. B}\ }%
  \textbf{\bibinfo {volume} {56}},\ \bibinfo {pages} {9552} (\bibinfo {year}
  {1997})%
  \bibAnnoteFile{NoStop}{sakurai1997}%
\bibitem{vanderhorst2001}%
  \BibitemOpen
  \bibfield{author}{%
  \bibinfo {author} {\bibfnamefont{J.-W.}\ \bibnamefont{van~der Horst}},
  \bibinfo {author} {\bibfnamefont{P.~A.}\ \bibnamefont{Bobbert}}, \bibinfo
  {author} {\bibfnamefont{M.~A.~J.}\ \bibnamefont{Michels}},\ and\ \bibinfo
  {author} {\bibfnamefont{H.}~\bibnamefont{B{\"a}ssler}},\ }%
  \bibfield{journal}{%
  \bibinfo {journal} {J. Chem. Phys.}\ }%
  \textbf{\bibinfo {volume} {114}},\ \bibinfo {pages} {6950} (\bibinfo {year}
  {2001})%
  \bibAnnoteFile{NoStop}{vanderhorst2001}%
\bibitem{eich2000}%
  \BibitemOpen
  \bibfield{author}{%
  \bibinfo {author} {\bibfnamefont{D.}~\bibnamefont{Eich}}, \bibinfo {author}
  {\bibfnamefont{D.}~\bibnamefont{Hubner}}, \bibinfo {author}
  {\bibfnamefont{R.}~\bibnamefont{Fink}}, \bibinfo {author}
  {\bibfnamefont{E.}~\bibnamefont{Umbach}}, \bibinfo {author}
  {\bibfnamefont{K.}~\bibnamefont{Ortner}}, \bibinfo {author}
  {\bibfnamefont{C.}~\bibnamefont{R. Becker}}, \bibinfo {author}
  {\bibfnamefont{G.}~\bibnamefont{Landwehr}},\ and\ \bibinfo {author}
  {\bibfnamefont{A.}~\bibnamefont{Fleszar}},\ }%
  \bibfield{journal}{%
  \bibinfo {journal} {Phys. Rev. B}\ }%
  \textbf{\bibinfo {volume} {61}},\ \bibinfo {pages} {12666} (\bibinfo {year}
  {2000})%
  \bibAnnoteFile{NoStop}{eich2000}%
\bibitem{huefner1991}%
  \BibitemOpen
  \bibfield{author}{%
  \bibinfo {author} {\bibfnamefont{S.}~\bibnamefont{Hufner}}, \bibinfo {author}
  {\bibfnamefont{P.}~\bibnamefont{Steiner}}, \bibinfo {author}
  {\bibfnamefont{I.}~\bibnamefont{Sander}}, \bibinfo {author}
  {\bibfnamefont{F.}~\bibnamefont{Reinert}}, \bibinfo {author}
  {\bibfnamefont{M.}~\bibnamefont{Schmitt}, \bibfnamefont{H~  and~Neumann}},\
  and\ \bibinfo {author} {\bibfnamefont{S.}~\bibnamefont{Witzel}},\ }%
  \bibfield{journal}{%
  \bibinfo {journal} {Sol. State Comm.}\ }%
  \textbf{\bibinfo {volume} {80}},\ \bibinfo {pages} {869} (\bibinfo {year}
  {1991})%
  \bibAnnoteFile{NoStop}{huefner1991}%
\bibitem{krause2008}%
  \BibitemOpen
  \bibfield{author}{%
  \bibinfo {author} {\bibfnamefont{S.}~\bibnamefont{Krause}}, \bibinfo {author}
  {\bibfnamefont{M.~B.}\ \bibnamefont{Casu}}, \bibinfo {author}
  {\bibfnamefont{A.}~\bibnamefont{Sch{\"o}ll}},\ and\ \bibinfo {author}
  {\bibfnamefont{E.}~\bibnamefont{Umbach}},\ }%
  \bibfield{journal}{%
  \bibinfo {journal} {New J. Phys.}\ }%
  \textbf{\bibinfo {volume} {10}},\ \bibinfo {pages} {085001} (\bibinfo {year}
  {2008})%
  \bibAnnoteFile{NoStop}{krause2008}%
\bibitem{scholz2009a}%
  \BibitemOpen
  \bibfield{author}{%
  \bibinfo {author} {\bibfnamefont{M.}~\bibnamefont{Scholz}}, \bibinfo {author}
  {\bibfnamefont{R.}~\bibnamefont{Schmidt}}, \bibinfo {author}
  {\bibfnamefont{S.}~\bibnamefont{Krause}}, \bibinfo {author}
  {\bibfnamefont{A.}~\bibnamefont{Sch{\"o}ll}}, \bibinfo {author}
  {\bibfnamefont{F.}~\bibnamefont{Reinert}},\ and\ \bibinfo {author}
  {\bibfnamefont{F.}~\bibnamefont{W{\"u}rthner}},\ }%
  \bibfield{journal}{%
  \bibinfo {journal} {Appl. Phys. A}\ }%
  \textbf{\bibinfo {volume} {95}},\ \bibinfo {pages} {285} (\bibinfo {year}
  {2009})%
  \bibAnnoteFile{NoStop}{scholz2009a}%
\bibitem{feng2005}%
  \BibitemOpen
  \bibfield{author}{%
  \bibinfo {author} {\bibfnamefont{D.}~\bibnamefont{Feng}}, \bibinfo {author}
  {\bibfnamefont{A.}~\bibnamefont{Caruso}}, \bibinfo {author}
  {\bibfnamefont{D.}~\bibnamefont{Schulz}}, \bibinfo {author}
  {\bibfnamefont{Y.}~\bibnamefont{Losovyj}},\ and\ \bibinfo {author}
  {\bibfnamefont{P.}~\bibnamefont{Dowben}},\ }%
  \bibfield{journal}{%
  \Doi{{10.1021/jp052712n}}{\bibinfo {journal} {J. Phys. Chem. B}}\ }%
  \textbf{\bibinfo {volume} {109}},\ \bibinfo {pages} {16382} (\bibinfo {year}
  {2005})%
  \bibAnnoteFile{NoStop}{feng2005}%
\bibitem{onsager1938}%
  \BibitemOpen
  \bibfield{author}{%
  \bibinfo {author} {\bibfnamefont{L.}~\bibnamefont{Onsager}},\ }%
  \bibfield{journal}{%
  \bibinfo {journal} {Phys. Rev.}\ }%
  \textbf{\bibinfo {volume} {54}},\ \bibinfo {pages} {554} (\bibinfo {year}
  {1938})%
  \bibAnnoteFile{NoStop}{onsager1938}%
\bibitem{braun1984}%
  \BibitemOpen
  \bibfield{author}{%
  \bibinfo {author} {\bibfnamefont{C.~L.}\ \bibnamefont{Braun}},\ }%
  \bibfield{journal}{%
  \bibinfo {journal} {J. Chem. Phys.}\ }%
  \textbf{\bibinfo {volume} {80}},\ \bibinfo {pages} {4157} (\bibinfo {year}
  {1984})%
  \bibAnnoteFile{NoStop}{braun1984}%
\bibitem{wojcik2009}%
  \BibitemOpen
  \bibfield{author}{%
  \bibinfo {author} {\bibfnamefont{M.}~\bibnamefont{Wojcik}}\ and\ \bibinfo
  {author} {\bibfnamefont{M.}~\bibnamefont{Tachiya}},\ }%
  \bibfield{journal}{%
  \bibinfo {journal} {J. Chem. Phys.}\ }%
  \textbf{\bibinfo {volume} {130}},\ \bibinfo {pages} {104107} (\bibinfo {year}
  {2009})%
  \bibAnnoteFile{NoStop}{wojcik2009}%
\bibitem{deibel2009a}%
  \BibitemOpen
  \bibfield{author}{%
  \bibinfo {author} {\bibfnamefont{C.}~\bibnamefont{Deibel}}, \bibinfo {author}
  {\bibfnamefont{T.}~\bibnamefont{Strobel}},\ and\ \bibinfo {author}
  {\bibfnamefont{V.}~\bibnamefont{Dyakonov}},\ }%
  \bibfield{journal}{%
  \bibinfo {journal} {Phys. Rev. Lett.}\ }%
  \textbf{\bibinfo {volume} {103}},\ \bibinfo {pages} {036402} (\bibinfo {year}
  {2009})%
  \bibAnnoteFile{NoStop}{deibel2009a}%
\bibitem{langevin1903}%
  \BibitemOpen
  \bibfield{author}{%
  \bibinfo {author} {\bibfnamefont{P.}~\bibnamefont{Langevin}},\ }%
  \bibfield{journal}{%
  \bibinfo {journal} {Ann. Chim. Phys.}\ }%
  \textbf{\bibinfo {volume} {28}},\ \bibinfo {pages} {433} (\bibinfo {year}
  {1903})%
  \bibAnnoteFile{NoStop}{langevin1903}%
\bibitem{shaw2008}%
  \BibitemOpen
  \bibfield{author}{%
  \bibinfo {author} {\bibfnamefont{P.~E.}\ \bibnamefont{Shaw}}, \bibinfo
  {author} {\bibfnamefont{A.}~\bibnamefont{Ruseckas}},\ and\ \bibinfo {author}
  {\bibfnamefont{I.~D.~W.}\ \bibnamefont{Samuel}},\ }%
  \bibfield{journal}{%
  \bibinfo {journal} {Adv. Mater.}\ }%
  \textbf{\bibinfo {volume} {20}},\ \bibinfo {pages} {3516} (\bibinfo {year}
  {2008})%
  \bibAnnoteFile{NoStop}{shaw2008}%
\bibitem{casu2007}%
  \BibitemOpen
  \bibfield{author}{%
  \bibinfo {author} {\bibfnamefont{M.~B.}\ \bibnamefont{Casu}}, \bibinfo
  {author} {\bibfnamefont{Y.}~\bibnamefont{Zou}}, \bibinfo {author}
  {\bibfnamefont{S.}~\bibnamefont{Kera}}, \bibinfo {author}
  {\bibfnamefont{D.}~\bibnamefont{Batchelor}}, \bibinfo {author}
  {\bibfnamefont{T.}~\bibnamefont{Schmidt}},\ and\ \bibinfo {author}
  {\bibfnamefont{E.}~\bibnamefont{Umbach}},\ }%
  \bibfield{journal}{%
  \bibinfo {journal} {Phys. Rev. B}\ }%
  \textbf{\bibinfo {volume} {76}},\ \bibinfo {pages} {{193311}} (\bibinfo
  {year} {2007})%
  \bibAnnoteFile{NoStop}{casu2007}%
\bibitem{sirringhaus1999}%
  \BibitemOpen
  \bibfield{author}{%
  \bibinfo {author} {\bibfnamefont{H.}~\bibnamefont{Sirringhaus}}, \bibinfo
  {author} {\bibfnamefont{P.~J.}\ \bibnamefont{Brown}}, \bibinfo {author}
  {\bibfnamefont{R.~H.}\ \bibnamefont{Friend}}, \bibinfo {author}
  {\bibfnamefont{M.~M.}\ \bibnamefont{Nielsen}}, \bibinfo {author}
  {\bibfnamefont{K.}~\bibnamefont{Bechgaard}}, \bibinfo {author}
  {\bibfnamefont{B.~M.~W.}\ \bibnamefont{Langeveld-Voss}}, \bibinfo {author}
  {\bibfnamefont{A.~J.~H.}\ \bibnamefont{Spiering}}, \bibinfo {author}
  {\bibfnamefont{R.~A.~J.}\ \bibnamefont{Janssen}}, \bibinfo {author}
  {\bibfnamefont{E.~W.}\ \bibnamefont{Meijer}}, \bibinfo {author}
  {\bibfnamefont{P.}~\bibnamefont{Herwig}},\ and\ \bibinfo {author}
  {\bibfnamefont{D.~M.}\ \bibnamefont{de~Leeuw}},\ }%
  \bibfield{journal}{%
  \bibinfo {journal} {Nature}\ }%
  \textbf{\bibinfo {volume} {401}},\ \bibinfo {pages} {685} (\bibinfo {year}
  {1999})%
  \bibAnnoteFile{NoStop}{sirringhaus1999}%
\bibitem{vanlaeke2006a}%
  \BibitemOpen
  \bibfield{author}{%
  \bibinfo {author} {\bibfnamefont{P.}~\bibnamefont{Vanlaeke}}, \bibinfo
  {author} {\bibfnamefont{A.}~\bibnamefont{Swinnen}}, \bibinfo {author}
  {\bibfnamefont{I.}~\bibnamefont{Haeldermans}}, \bibinfo {author}
  {\bibfnamefont{G.}~\bibnamefont{Vanhoyland}}, \bibinfo {author}
  {\bibfnamefont{T.}~\bibnamefont{Aernouts}}, \bibinfo {author}
  {\bibfnamefont{D.}~\bibnamefont{Cheyns}}, \bibinfo {author}
  {\bibfnamefont{C.}~\bibnamefont{Deibel}}, \bibinfo {author}
  {\bibfnamefont{J.}~\bibnamefont{D'Haen}}, \bibinfo {author}
  {\bibfnamefont{P.}~\bibnamefont{Heremans}}, \bibinfo {author}
  {\bibfnamefont{J.}~\bibnamefont{Poortmans}},\ and\ \bibinfo {author}
  {\bibfnamefont{J.~V.}\ \bibnamefont{Manca}},\ }%
  \bibfield{journal}{%
  \bibinfo {journal} {Sol. Ener. Mat. Sol. Cells}\ }%
  \textbf{\bibinfo {volume} {90}},\ \bibinfo {pages} {2150} (\bibinfo {year}
  {2006})%
  \bibAnnoteFile{NoStop}{vanlaeke2006a}%
\bibitem{osterbacka2000}%
  \BibitemOpen
  \bibfield{author}{%
  \bibinfo {author} {\bibfnamefont{R.}~\bibnamefont{{\"O}sterbacka}}, \bibinfo
  {author} {\bibfnamefont{C.~P.}\ \bibnamefont{An}}, \bibinfo {author}
  {\bibfnamefont{X.~M.}\ \bibnamefont{Jiang}},\ and\ \bibinfo {author}
  {\bibfnamefont{Z.~V.}\ \bibnamefont{Vardeny}},\ }%
  \bibfield{journal}{%
  \bibinfo {journal} {Science}\ }%
  \textbf{\bibinfo {volume} {287}},\ \bibinfo {pages} {839} (\bibinfo {year}
  {2000})%
  \bibAnnoteFile{NoStop}{osterbacka2000}%
\end{thebibliography}
\end{document}